# A Heuristic Routing Mechanism Using a New Addressing Scheme

M. Ravanbakhsh[1,2], Y. Abbasi-Yadkori[1,2], M. Abbaspour[1,3], H. Sarbazi-Azad[1,2]

[1]*IPM School of Computer Science, Tehran, Iran*
[2]*Department of Computer Engineering, Sharif University of Technology, Tehran, Iran*
[3]*Department of Electrical and Computer Engineering, Shahid-Beheshti University, Tehran, Iran*
y_abbasi@ce.sharif.edu, ravanbakhsh@ce.sharif.edu, maghsoud@ipm.ir, azad@{ipm.ir, sharif.edu}

**Abstract.** *Current methods of routing are based on network information in the form of routing tables, in which routing protocols determine how to update the tables according to the network changes. Despite the variability of data in routing tables, node addresses are constant. In this paper, we first introduce the new concept of variable addresses, which results in a novel framework to cope with routing problems using heuristic solutions. Then we propose a heuristic routing mechanism based on the application of genes for determination of network addresses in a variable address network and describe how this method flexibly solves different problems and induces new ideas in providing integral solutions for variety of problems. The case of ad-hoc networks is where simulation results are more supportive and original solutions have been proposed for issues like mobility.*

## 1. Introduction

Addressing and routing are two of the most important elements in modern network infrastructures. Recently, new problems like congestion in spite of disengaged paths [1], lack of a fair and perfect pricing mechanism [5], multicast and grouping overheads [7] and newborn problems of ad-hoc networks, have challenged the networking industry. There is a wide spread belief that because of an inherent unity between these problems in nature, providing particular solutions to each of them might prevent the others from becoming practical. In other words, there is a feeling of necessity for a unified solution for these problems, a solution that may act more flexible in confronting new concerns and introduction of new concepts. Previously, however, there have been some works, though considering simplified conditions, to achieve a general solution to the problem of routing efficiency and pricing using an approach inspired by the game theory [24].

In this paper, we first introduce an addressing scheme using variable addresses. Then, we propose an integral solution to some of prevalent problems including some in the area of ad-hoc networks like mobility and power-management, using a heuristic method named *genetic routing* based on the concept of variable addresses. Although the proposed solutions are not fully descriptive, they help clarifying the idea. As the topic of this article implies, rather than trying to describe a detailed algorithm, our attempt is to explain a mechanism of dealing with such issues, so there might be various implementations with different effectiveness for a particular usage.

The rest of the paper is organized as follows. Section 2 introduces the proposed addressing scheme. In section 3, a heuristic routing based on genetic algorithms is introduced. Section 4 discusses the routing in mobile ad-hoc networks. Some simulation results are reported in section 5, followed by section 6 which concludes the paper.

## 2. A network of nodes with variable addresses

In this section, first we describe an overall sketch of the addressing mechanism and then give a more precise definition in mathematical notation.

Consider a network in which every node has a long enough variable address to reflect the topology and the traffic of the time for that node. The exact meaning of the term *reflect* becomes clearer in the rest of the text. However, it is not wrong to think of node address as a replacement for keeping and exchanging the routing table's data. Then, routing is defined as follows: every node upon receiving a packet with a particular destination, applies a comparison function, with every neighboring node address and the destination address being its arguments, to determine where to forward the packet. For example if the comparison function is *similarity*, the routing method forwards the packet to a neighboring node which is the most similar to the destination address.

Let $G(V, E)$ be the network graph, where $V = \{v_i\}$ is the set of network nodes, and $E = \{e(v_i, v_j) | v_i, v_j \in V\}$ represents the set of instantaneous links costs. The link cost between two nodes $u$ and $v$ is assumed to be $\infty$ when no direct connection exists between them. Please note that we consider $G$ as a snapshot of the network topology, in which the links costs represent the network traffic in that particular moment. Our aim is to assign an address $a(v_i)$ to each node $v_i$ at the moment under consideration, where

$$a : V \to F,$$

$$F = \{f : \{1,...,address\_length\} \to \{0,...,(Base-1)\}\},$$

Here, $F$ denotes a set of all possible functions as defined above. Actually an address $a(v_i)$ denotes a sequence of specified length *address_length* that consists of integer numbers in the given base *Base*. The employed set notation leads to the simplicity of future definitions.

Both of the parameters *Base* and *address_length* are network parameters. A network parameter is a parameter that should be defined according to the network conditions including size, topology and routing algorithm in use (as previously mentioned, we can have different implementations).

When forwarding a packet, we use the evaluation function *eval* to determine the priority of the neighboring nodes, where

$$eval : F \times F \to \Re,$$
$$eval(f_i, f_j) = \overline{(F_i \cap F_j)},$$

where $F_i$ and $F_j$ are the set representations of $f_i$ and $f_j$ (as a binary relation), and two parallel lines above a set means cardinality of that set. Please note that one could implement the function *eval* in any other way. We have used the intersection operation, because actually it is the one we use in the simplest implementation of genetic routing.

We also define a complete graph $C(V, E')$ as the *Distance-Graph*, in which $V$ is the same as in $G$, and

$$E' = \{e'(v_i, v_j) = eval(a(v_i), a(v_j))\}.$$

Now, let us have a look at the problem from the opposite side. Having the value of the evaluation function for every two nodes, we may try to find the proper address for each node. These given values of the evaluation function are determined by a routing protocol. In fact, a routing protocol shall be defined as a method of mapping $G \to C$ in a distributed manner.

Now, the question is, having the graph $C$, how we can find a consistent address assignment function '$a$' if one exists. And what characteristics make an assignment feasible for a distance-graph.

This question defines a *Constraint Satisfaction Problem*. Although the problem seems to be NP-complete, it has a solution with a run time complexity of $O(n)$, where $n$ is the number of nodes in the network. It is because the constraint network is strongly n-consistent if some weak conditions held. Thus, there is a solution without backtracking for any variable ordering (see *Appendix A*). Using the proposed notation, contemporary routing protocols correspond to some relatively simple and straightforward assignments of the evaluation function. Here, by *simple* we mean there is the possibility of more complex assignments by which we may have a better consideration of the traffic and topology. Also, introducing a distributed implementation of the assignment algorithm still remains a challenging problem. For example, the *OSPF* routing protocol, that uses time as the metric, can be simply defined as $e(v_i, v_j)$, i.e. the length of a shortest path between $v_i$ and $v_j$'.

Since similarity is a symmetric function, it seems that the proposed addressing scheme is not able to cause proper traffic effects in addresses when dealing with asymmetric traffic patterns. However, the evaluation function can be an asymmetric function in order to reflect asymmetric traffics.

The most crucial problem with variable address mapping is the issue of variability, which result in an important question: "How can one node address another node with a variable address?" In contrary to its difficult appearance, it's easy to solve because:

- Address invariability is only necessary for end-nodes not for the intermediate ones (of course this is not true in the case of ad-hoc networks).
- The address variation is only the result of traffic change. On the other hand, the address of a node is a representation of the topology as well, which is supposed to be constant (for MANETS this is not the case and the solution is discussed in section 4-1).
- Tree-like structure of topology is common at the borders of networks. On the other hand, in a tree-like topology, the path remains the same, whatever the traffic is. Thus, the border nodes, including the end-nodes, can easily just reflect the topology.
- In the worst case, the end-nodes can have permanent addresses that only reflect the topology, so there would be more burdens on the middle way routers to bring about flexibility.

## 3. The proposed routing algorithm

The *genetic routing* algorithm is based on variable address assignment, which exploits the concept of 'gene' in a similar way to its expected meaning. Although the algorithm, in its essence, is not like other genetic algorithms, for the sake of simplicity we call it genetic routing. The following example demonstrates how a heuristic method can be merged with the concept of variable addresses, and how it can contribute to the emergence of a new routing algorithm.

As we defined previously, in a variable address network, every node has a fixed-length variable address, which is the output of the function '$a$' having that node as the argument. Genes are responsible for making similarities in nodes by copying and pasting address segments. Genes are small packets (at most tens of bytes) that network nodes can distinguish them from data packets and treat them differently. Consider the graph $G(V,E)$ as the network graph in which address mapping '$a$' is a *random* assignment of $f$ to $v$, where $f \in F$ and $v \in V$.

If a gene $g$ is generated in the node $v_i$, then it is a subset of $F_i$. After transmission of $g$ to $v_j$, the node $v_j$ applies that gene to itself, and then we have $g \subseteq F_j$. We say the gene has *affected* the node or has made the similarity. Suppose every node has a specific amount of genes. These genes are distributed among neighbors. Each neighbor after applying gene to itself (by applying, we mean affecting the node address by the gene), would carry out one of the following actions in a probabilistic manner:
- It would alter the gene according to its own address and then send it back to the sender, or
- (with the probability of distribution-percentile as a network parameter) send it unchanged to one of its other neighbors.

An approved suggestion for the evaluation function is given in Appendix *A*.

There are some issues to be discussed, not because of their importance in this implementation, but because they show how flexible this approach is in giving an original solution when confronting a problem.

**Reflecting the traffic:** Node addresses during the convergence time somehow indicate the traffic at that time. During the last moments of convergence, packets can partially find their route, since there is a convergence of node addresses although not indicating the final path. On the other hand, consider reciprocation of genes on that link, in a particular duration. As said before, a node may alter a gene and send it back to its previous sender; this would cause the reciprocation of some genes between neighbors. In a congested link, there would be less gene exchange in the same interval because genes should spend much of their time in the queue, while on a disengaged link, more similarity might be made between two neighbors as a result of gene exchange. But since there is no consideration for traffic asymmetry, this change will not be as effective as expected. Consider a link which is congested just in one way; the resulting similarity of the addresses would not affect the addresses in an asymmetric manner to show which way is congested and which is disengaged, but it works well when the traffic is nearly symmetric.

**Elimination of genes**: Since we need to eliminate genes from the network when enough convergence is obtained, we accompany every gene with a mask array where each cell corresponds to a number in the gene's address. More precisely

$$mask(g) : g \rightarrow \{active, passive\},$$

determines whether the corresponding number in the gene is active or passive. When passive, a number can not participate in the copy and paste process. In each paste action, one mask bit is randomly selected and it becomes false if it is not false already. When all bits of a mask are false, a gene is no more effective and a node can neglect it. (The relation between lifetime and gene-length is calculated in Appendix B).

### 3.1 Advantages of the proposed routing

Here, we describe some merits of the proposed genetic routing algorithm.

1. Its simplicity as
   - It is obvious that the routing process is solely based on node addresses without any peripheral information of the network.
   - Forwarding process is simply a group bit comparison operation and if *Base* is chosen to be a power of two, a simple hardware implementation is possible.
   - Just like the forwarding process, interpretation and update of genes, is quite simple and makes fast hardware implementations possible (actually we should just copy and paste groups of bits with fixed length; the most complex part is to behave randomly).
2. After convergence, a loop free routing and correct delivery is guaranteed, which is in contrary to Distance-Vector routing protocols like RIP, where an extra loop elimination mechanism is essential.
3. In contrary to the other protocols (see [8]), the drop of protocol messages, not only wont cause any problem but it is a merit since it reflects congestion. Every time we have a gene dropped in a queue as a result of congestion, it would lessen the amount of similarity between those two neighbors which is in favor of this routing protocol.
4. Each end-node has a good estimate of its distance to the other end-node, which turns to be useful in higher layer protocols for reliable transfer, and also in selecting between different replications of data or a service provided by different hosts.
5. *Pricing*: Since the difference between the current and the destination node addresses of a packet is a good estimate of the network resources it uses (this is the case specially when a variable address routing protocol takes traffic more exactly and asymmetrically into account, in which congested links would have their actual price) there is a possibility of exploiting a packet based charging mechanism (see [5]) which uses a reverse path charging. With this schema, a correct value of each packet is taken into account and each intermediate router also receives as much as it provides resources. On the other hand, pricing would act as a motivator to reduce the congestion and balance the network traffic, which was impossible to balance without cooperation of end-nodes. One example of such application is discussed by Gibbens and Kelly in [25]. Generally, using this method of pricing, there is a possibility for the existence of a Nash-equilibrium in network resource usage, which in turn will probably result in a minimum latency for all agents [26].
6. Applying different kinds of *policies* is possible for each node in two forms:
   - As an indicator of activity in the routing process, a node, by the amount of genes it produces, can decide to what extent it should influence other nodes, and by the amount of receiving genes it applies to itself, can decide to what extent it should be affected.

- An intermediate node can increase or decrease the amount of the traffic it receives from its neighbor x that should be transmitted to the neighbor y, by increasing or decreasing its own similarity to their common similarity through deciding on the amount of received genes from one of them that should be passed to the other one during the convergence time. However, it is possible to adjust those similarity characteristics in any arbitrary time as discussed in Appendix A. consider the node *v* which has two neighbors *x* and *y*, and let $S_{x,y} = F_x \cap F_y$. The cardinality of $S^v_{x,y} = S_{x,y} \cap F_v$ influences the amount of traffic between x and y which passes through node v.

7. There are variations of this routing mechanism that use asymmetric evaluation functions instead of intersection. They search for particular patterns in each node address and use that pattern to determine which node they should forward the packet to. A pattern may be a simple one based on the recurrence of a particular integer in an address, or any other more complex ones. These variations are capable of implementing multicast and anycast in their optimal (see [7, 14]) and stateless form, using common patterns in group addresses. By stateless we mean there is no need for keeping group states in intermediate routers as is needed for current protocols.
8. In case of link or node failure, which in turn is similar to some concurrent link failure, the corresponding packets can still be delivered to an available node which is the most similar to the destination; current prevalent protocols lack such capability [10]. Although this increases the packet delay, it also prevents from packet loss. To resolve the problem permanently, two disconnected nodes shall remove similarities from their addresses and send genes, which are copied from their changed segments, to their other neighbors to affect the neighbors according to such disconnection. Somehow it is an attempt to eliminate the prior similarities. The other solution is to deliver the packets through available paths and wait for the next global convergence to overcome the lack of optimality.
9. The concept of genes provides the possibility of influencing local decisions by global conditions, which is a necessity for a forth generation routing protocol [3, 13].

## 4. Genetic routing in mobile ad-hoc networks

The genetic routing method works well particularly for ad-hoc networks, in which we have locality in connections, in contrast to wired ones in which there is no necessity for locality of a connection. By locality in connections we mean some kind of geographical factors in deciding the connectivity of two nodes which finally causes a semi-planar graph (see [15] for discussion of modeling using such geography-based topologies). For more information on MANETs see [9, 11, 16].

Flexibility of the concept of genes in inclusion of new functionalities is evident in confronting the issues discussed below.

### 4.1 Mobility

Handling mobility in MANETs is a great challenge in designing new routing protocols. When a mobile node changes its place how it should be assigned a new address and how the other nodes can route to it, just having its old address? These are the main questions.

Currently we have two suggestions for resolving the issue of mobility using variations of genetic routing. The first one would be considered as the selected one because of its simplicity and performance which is also approved by simulations, but the second one has its own virtues in grouping.

o The mechanism is simple: a *periodic global converge*. The period of converge is determined by the amount of mobility in the network. A highly mobile network in which mobile nodes exit one region and join another one rapidly needs shorter periods while a more static network in which mobile nodes have lower speeds would be satisfied with longer convergence periods. But how a particular node is supposed to find another one when after a periodic global converge, nodes would have new addresses. The answer is: nodes will exploit their old addresses until the new convergence become complete, and then they try to switch to their new address while still keeping their own and their neighbor's old addresses. In this manner, every node would have a list of say *K* previous periodic global convergence addresses of its own and its neighbors. The case is trivial while nodes are in communication during the address switching since they would inform each other of their new address before switching. Now consider a node which wants to route to another one, but the latest address it has from that node is the address which belongs to *K'* period of global convergence before. This node still can route to the destination in order to achieve its new address if K' < K, and then continue the communication using the new address. Consider two scenarios, in the first one the destination node has not moved since K' last period, and in the second one it has moved.

- For the *first scenario*, the method is easy. Produce a packet with the old destination address, but also label the packet to determine to which periodic convergence it belongs, say K' periods before now. Since every node has a record of its own and its neighbors address in K' period of global convergence, every node can easily check to see which of them have stayed in the same region since then and select the most similar one according to the old address and forward the packet correctly. If there is a connection available between static nodes, since then the packet would route correctly. And they can achieve their new address to communicate using it.

- For the *second scenario*, the solution is a little more algorithmic. Every mobile node after a movement, to stay reachable from other agents, should register its new address in its old neighbors. In this case, another node having an old address can find its new address by sending a packet with old address and label of age (to which period of global convergence before it belongs) to destination. Then the mobile node old neighbors would inform the sender of new address, and everything goes on with new address.
- We should notice that the necessity of finding any particular node having its previous address is not always the case. In many situations, having the address of some particular nodes is enough; this is the case with gateways and master nodes. In these cases, we can merely use the permanent address for those nodes. Still network addresses would converge and work properly while number of such nodes is small.

o There is also an alternative approach which is based on gene patterns in forwarding mechanism
o rather than merely counting similarities. In those methods, mobility can be implemented by producing genes carrying special patterns to new region in the case of arrival and spreading genes for removing available patterns from leaving area before departure.

## 4.2 Power management

Managing the overall network power to keep the network graph connected as long as possible, so using low power agents less than the high power ones in the process of routing [12, 18, 20], using lower cost paths with the power as metric [2], and determining the efficient power range of a node are important issues in MANETs. We won't solve them all, though the first and the second one are in opposition. As was mentioned in applying policies, one node can easily decide its involvement in routing, even between two particular neighbors which here may be in long geographical distance. So it's up to every node to manage its own power consumption regarding the other nodes in its neighborhood, since genes may act as an indication of a node's power too. Every node can decide not to participate in routing between too active neighbors or neighbors locating in relatively far distance to keep its own power in desired level or keep the total network connected as long as possible. In another way a node can reduce its overall involvement in routing process regardless of special neighbors in a same way of applying policies (discussed in 3-1.6).

## 4.3 Address assignment

### 4.3.1 Initial assignment

In MANETs, dynamic and fast address assignment, with low overhead is an important issue (see [17, 21, 22]). In genetic routing, setting an address for a newly joined node is easy; two suggestions are:

o Wait until the next periodic global converge; then as a result of convergence, newly joined node would get a new address.

o If the node doesn't want to wait until the next convergence, it can get neighbors address and assign a mixture of their address manually to itself. This address although not precisely assigned, would work correctly, since the assumption is that the packets would route correctly to neighbors.

### 4.3.2 Merging and partitioning

Both of these functionalities are easy to achieve using the method of periodic convergence. We don't describe them as the idea is obvious.

## 4.5 Grouping

Grouping and routing according to the common patterns is possible as mentioned before. In addition to providing stateless optimum multicast and anycast, this schema can provide facilities for exploiting dominating sets [4] in order to have lower power consumption noticing the advantages of genes in algorithms for finding dominating sets. For an example, a node can easily determine if its two neighbors are directly connected or not, calculating their similarity without any message exchange. Genes also provide facilities for selecting Virtual Backbones [19], since a backbone link should have a reasonable bandwidth and also should have stayed static long enough, for both of which we have genetic metrics in genetic routing.

## 5. Simulations results

We simulated a simple implementation of proposed genetic routing for both ad-hoc and wired networks of 10, 20 and 50 nodes for different topologies for each network size. Results for varieties of network parameters are shown in table 1.

In this table, *c.t* is the average convergence time, with all network links having 10 KBytes of bandwidth. Considering approximately linear relation of c.t with the inverse of bandwidth, because of negligible processing time required for genes, **c.t** can be improved so much in networks with high bandwidth. Efficiency *(eff)* is measured in comparison with SPF average network delay.

***n.g*** is the average amount of genes every node has. ***n.w*** is the average network width (maximum over every pair of minimum distance between two nodes, measured in number of hops). ***g.l*** is the gene length. ***a.l*** is address length in number of digits in Base ***b*** and finally ***n.s*** is the network size in number of nodes. As you may have noticed, there is a tradeoff between various desired characteristics.

**Table 1: Converge time and efficiency for wired network.**

| n.s | c.t (ms) | n.w | Eff | g.l | b | a.l | n.g |
|---|---|---|---|---|---|---|---|
| 10 | 8 | 3.2 | 1.350 | 1 | 16 | 50 | 2.5 |
| 10 | 10 | 3.2 | 1.000 | 1 | 16 | 50 | 3 |
| 10 | 11 | 3.2 | 1.000 | 1 | 16 | 50 | 3.5 |
| 20 | 650 | 6.2 | 1.005 | 10 | 32 | 50 | 7 |
| 20 | 420 | 6.2 | 1.002 | 8 | 32 | 50 | 7 |
| 20 | 369 | 6.2 | 1.010 | 8 | 32 | 50 | 6.2 |
| 20 | 295 | 6.2 | 1.014 | 7 | 32 | 50 | 6.2 |
| 20 | 260 | 6.2 | 1.092 | 7 | 32 | 50 | 5.0 |
| 50 | 429 | 6.8 | 1.000 | 7 | 32 | 100 | 6.0 |
| 50 | 360 | 6.8 | 1.046 | 7 | 32 | 100 | 5.28 |
| 50 | 300 | 6.8 | 1.000 | 6 | 32 | 100 | 5.28 |
| 50 | 2812 | 8 | 1.022 | 10 | 32 | 100 | 32 |
| 50 | 1474 | 8 | 1.105 | 5 | 32 | 100 | 43 |
| 50 | 1211 | 8 | 1.037 | 4 | 32 | 100 | 46 |

Protocol overhead is an important parameter in ad-hoc routing protocols [23]. The measurement of protocol overhead during the convergence time has also been appended for ad-hoc networks.

Here, ***p.o*** is the protocol overhead per link during the convergence time (***c.t***) and ***n.t*** shows the density of nodes, dividing networks to sparse (***s***) and dense (***d***) types.

## 5. Conclusion

The quick convergence and little protocol overhead for small networks are obvious from tables 1 and 2. It is enough to say that for small networks the convergence is possible even with genes of length one, which means the convergence occurs in one step, in comparison with RIP which needs *n* steps to converge (*n* is the number of nodes).

Another result of simulations is that genetic routing mechanism might be highly advisable in networks with locality in connections, like ad-hoc networks. Although convergence time increases fast as network size grows, protocol overhead grows reasonably. In the proposed method of periodic convergence for MANETs, by extending the convergence time to convergence period, the amount of protocol overhead would come even lower. Having all these results with the merits mentioned in 3.1, reveal that the proposed mechanism of routing is really attractive.

**Table2**: Convergence time and protocol overhead for genetic routing in ad-hoc networks.

| n.s | n.t | c.t (ms) | n.w | p.o/link (Bytes/s) | g.l | b | a.l | n.g |
|---|---|---|---|---|---|---|---|---|
| 10 | s | 12 | 4.3 | 277 | 1 | 16 | 50 | 5 |
| 10 | s | 4 | 4.3 | 333 | 1 | 16 | 50 | 2 |
| 10 | s | 9 | 4.3 | 296 | 1 | 16 | 50 | 4 |
| 10 | s | 13 | 4.3 | 307 | 1 | 16 | 50 | 6 |
| 10 | s | 10 | 4.6 | 357 | 1 | 16 | 50 | 5 |
| 10 | s | 9 | 4.6 | 238 | 1 | 16 | 50 | 3 |
| 10 | s | 9 | 4.6 | 317 | 1 | 16 | 50 | 4 |
| 20 | s | 52 | 7.0 | 720 | 2 | 32 | 50 | 10 |
| 20 | s | 49 | 7.0 | 942 | 3 | 32 | 50 | 5 |
| 20 | d | 11 | 6.4 | 275 | 1 | 32 | 50 | 5 |
| 20 | d | 23 | 6.4 | 263 | 1 | 32 | 50 | 10 |
| 50 | s | 771 | 17.0 | 970 | 4 | 32 | 100 | 50 |
| 50 | d | 540 | 8.0 | 439 | 3 | 32 | 100 | 50 |
| 50 | d | 215 | 8.0 | 444 | 3 | 32 | 100 | 20 |
| 50 | d | 545 | 8 | 503 | 4 | 32 | 100 | 30 |

Future work in this line of research can: a) search for a distributed evaluation function to optimize the variable address network's routing mechanism, b) look for possibility of finding a Nash-equilibrium which also maximizes the network throughput based on the proposed schema of reverse path charging, c) simulate alternative implementations of genetic routing with different evaluation functions, d) simulate efficiency of suggested optimum stateless multicasting and anycasting, e) further attempt in integrating ad-hoc issues in the model and verification of their functionality, and f) extensively run simulation experiments for evaluation of efficiency of proposed mechanism for power management.

# Appendix A

There can be an assignment of address to a node *(N)*, with the constraint of having predetermined similarity ($k_i$) with each of its neighbors *($V_i$)*; however, some weak conditions are necessary to be held. Having a mapping $a(V_i)_{v_i \neq N}$, we want to determine an address *a(N)* with the following constraint

$$\forall_{1 \leq i \leq n} \overline{a(N) \cap a(v_i)} = k_i \quad (1)$$

where *i* is counting the neighbors.

If $\sum_{i=1}^{n} k_i \leq address - length$, finding a solution satisfying equation (1) is trivial, as we can dedicate a piece of address for similarity with a special neighbor.

However, when $\sum_{i=1}^{n} k_i > address - length$, then the necessary weak condition is

$$\sum_{i=1}^{n} \sum_{j=1}^{n} \overline{a(v_i) \cap a(v_j)} \geq \sum_{i=1}^{n} k_i - address\text{-}length \quad (2)$$

The reason is because we are trying to assign an address to *N* using common digits in neighbors. But also there are conditions in the network which may guarantee the validity of required condition (2). In a network with (*Base < n*), which is usually the case, we have $\sum_{i=1}^{n} \sum_{j=1, j \neq i}^{n} \overline{a(v_i) \cap a(v_j)} \geq [address\text{-}length \times (n - Base)]$.

It is because, regarding an address digit and according to the pigeon-hole principle until $n \leq Base$, there could be different address digits for different nodes; so if *n* become bigger than *Base* for each node, there should be at least one repetition of that digit. So, the overall amount of similarity is at least the sum over all (*address-length*) values. Then, it is sufficient to have:

$$address\text{-}length \times (n - Base) \geq \sum_{i=1}^{n} k_i - address\text{-}length$$

This condition is satisfied for a large enough *n*, since *Base* and address length are constant and $k_i$ is always smaller than address-length. So, the inequality holds for large enough values of *n*. In this case, assignment is easy. First, we assign according to common digits of neighbors and the remaining digits from individual neighbors to satisfy condition (1).

The overall procedure of mapping is easy now. We can start with a node and assign it an arbitrary address. The next one should be assigned in a way so that the constraint of their predetermined similarity be satisfied. The next one should have an address consistent with two previous ones, and so on.

Since we can have an assignment in the case of (*n*-1) constraints all assignments can be done without backtracking.

## Appendix B

Although the model has a stochastic behavior, there are some facts, which can be estimated. The most important one is the evaluation function *eval*. There is a mathematical relation between network parameters as well, which shall be exploited to avoid or shorten the necessary process of trial and error. This trial and error process is necessary for adjusting parameters for best convergence and lowest overhead. One of them is the relation between gene-length(GL) and its overall activity

$$\Sigma = GL + \log_{\left(1-\frac{1}{GL}\right)}^{\frac{1}{GL}} \qquad (4)$$

where $\Sigma$ is the number of gene's reciprocations.(We don't prove these relations, since they're not that much important) One other important parameter is the relation between two neighbors' similarity and the number of gene exchange between them. One fact is, if (address-length(AL) >> gene-length) there is no difference between k exchange of a gene with length one and one exchange with length k then, having initial similarity $S_1$ after one gene exchange

$$S_2 = \sum_{i=0}^{GL-2}\left(1-\frac{1}{AL}\right)^i + S_1\left(1-\frac{1}{AL}\right)^{GL-1}$$

Assuming average initial similarity of $\frac{address\text{-}length}{Base}$, consequent similarities are computable, knowing number of initial genes for each node, and using (4).

For this method of address mapping, evaluation function can be expressed as:

$$eval(v_i, v_j) = \sum_{\lambda} \frac{\lambda_{i,j}}{\kappa \times \delta^{length(\lambda_{i,j})}}$$

in which $\lambda_{i,j}$ is the amount of flow in a segment of maximum flow (flow of augmenting path) between $v_i$ and $v_j$, $\kappa$ is a constant and $\delta$ is the average out-degree of network graph, and length ($\lambda_{i,j}$) is number of hops that flow segment covers. Because value of $\kappa$ is highly dependent of the degree of network convergence, precise verification of this formula was not possible, although for the same size networks of 10, 20 nodes the formula was a good estimation for best cases of converge. Intuitive support of this formula is:

We know that the probability of a gene to get to a node, $\ell$ nodes away, in a network with average out-degree of $\delta$, is $1/\delta^\ell$. This explains the denominator of the relation.

Also the effects of genes are dependent of the quality of flow rather than the number of paths, since two paths may share a same portion but for flow this is not the case. The more flow available between two nodes, the better genes can act in causing similarity between them; however the distance in each path remains important.